\documentclass{PoS}

\usepackage{epstopdf}

\title{Hints of BSM physics in the SM effective potential}

\ShortTitle{BSM in the SM effective potential}

\author{\speaker{Zygmunt Lalak}\\
        Institute of Theoretical Physics, Faculty of Physics, University of Warsaw ul. Pasteura 5, 02-093 Warsaw, Poland\\ 
        E-mail: \email{Zygmunt.Lalak@fuw.edu.pl}}

\author{Marek Lewicki\\
        Institute of Theoretical Physics, Faculty of Physics, University of Warsaw ul. Pasteura 5, 02-093 Warsaw, Poland\\
        E-mail: \email{Marek.Lewicki@fuw.edu.pl}}
        
\author{Pawe\l{} Olszewski\\
        Institute of Theoretical Physics, Faculty of Physics, University of Warsaw ul. Pasteura 5, 02-093 Warsaw, Poland\\
        E-mail: \email{Pawel.Olszewski@fuw.edu.pl}}

\abstract{Investigation of the structure of the Standard Model effective potential at very large field strengths opens a window towards new phenomena and can reveal properties of the UV completion of the SM. The map of the lifetimes of the vacua of the SM enhanced by nonrenormalizable scalar couplings has been compiled to show how new interactions modify stability of the electroweak vacuum. Whereas it is possible to stabilize the SM by adding Planck scale suppressed interactions and taking into account running of the new couplings, the generic effect is shortening the lifetime and hence further destabilisation of the SM electroweak vacuum. Absolute stability can be achieved by lowering the suppression scale of higher order operators while picking up such combinations of new couplings, which do not generate new deep minima in the potential. We discuss the issue of gauge dependence of the perturbative determination of the tunnelling rate and show how this rate can be made gauge independent at the leading nontrivial order of the RGE improved effective action.}

\FullConference{Proceedings of the Corfu Summer Institute 2014 "School and Workshops on Elementary Particle Physics and Gravity",\\
		3-21 September 2014\\
		Corfu, Greece }

\begin{document}
\section{Introduction}
The discovery of the 125 GeV scalar particle, which in the light of available data can be identified with the Standard Model Higgs boson,  and absence of experimental signature of any new physical state in the LHC experiments makes it important to search for possible windows towards new phenomena within the Stadard Model itself. One of the possible windows is the investigation of the structure of the effective potential in the Standard Model which has been the subject of considerable activity  \cite{Coleman,Buttazzo:2013uya,Degrassi,Ellis:2009tp,Espinosa,Casas:2000mn,Casas:1996aq,Casas:1994qy,Sher,Branchina:2013jra}. 


The question about stability of the SM vacuum in the presence of ultraviolet completions at or below the planck scale is the central point of this note.  In the paper \cite{Branchina:2013jra} higher order operators have been added to the scalar potential of the neutral Higgs field. The operators were assumed to be suppressed by suitable powers of the Planck scale and for sensible values of the new couplings they were found to modify significantly the behaviour of the potential near the Planck scale. 
%
In the paper \cite{Lalak:2014qua} the question has been studied further, in particular a complete map of the vacua in the SM extended by nonrenormalisable scalar couplings, taking into account the running of the new couplings, have been produced. Also going beyond the standard simplifying assumptions taken when calculating the lifetime of the metastable vacuum and instead using a completely numerical approach. 

However, we begin with the discussion of the issue of gauge dependence of the perturbative determination of the tunnelling rate and argue how this rate can be made gauge independent at the leading nontrivial order of the RGE improved effective action.

\section{Gauge dependence of tunnelling rate}
It is well known that the effective potential and in general the effective action are gauge-dependent objects. However, using them properly one can draw gauge independent physical conclusions. In particular the statement about the spontaneous breaking of gauge symmetry is gauge invariant \cite{Nielsen:1975fs}.
The gauge invariant "observables" are the values of the effective potential at the extrema, and the tunneling rate between different minima. 
The position of the minima is gauge dependent, the values of the classical field $\phi_c$ at arbitrary points are gauge dependent as well. 
However, when one computes the SM effective potential in a straightforward manner (say naively), nothing looks gauge independent - neither the value of the effective potential at the extrema (see \cite{DiLuzio:2014bua}) nor the tunneling rate. In order to obtain gauge invariant quantities one must perform appropriate expansion of the effective action. The relevant strategy to demonstrate vacuum lifetime gauge independence has been discussed in \cite{Metaxas:1995ab}, and applied there to the case of scalar electrodynamics. 
Very similar treatment was recently used also in case of the Standard Model in order to show the gauge independence of values of the potential at its extrema in \cite{Andreassen:2014eha,Andreassen:2014gha}. There the authors point out two main problems one faces while using the approach of \cite{Metaxas:1995ab} to the Standard Model. Firstly, RGE improved renormalisation of the field strength $\Gamma=\int \gamma \ln \mu$ induces gauge dependant correction at all orders of perturbation theory. Secondly, the expansion used in \cite{Metaxas:1995ab} assumes $\lambda \propto \hbar e^4$ (where $e$ is the gauge coupling constant) which is also violated by RGE resummation. In this note we shall briefly discuss possible solution to these problems in case of the Standard Model. More exhaustive discussion will appear elsewhere. 
In order to illustrate the gauge dependence induced by field strength renormalisation it is enough to consider tree level potential improved with one-loop renormalisation group equations. For electroweak vacuum stability consideration this implies a very simple quartic potential
\begin{equation}\label{simplepotential}
V=\frac{\lambda(\mu)}{4}e^{4\Gamma(\mu)}h^4
\end{equation}
where $\Gamma(\mu)=\int_{M_t}^{\mu} \gamma(\bar{\mu}) \ln \bar{\mu}$. The next step in the standard procedure is to set $\mu=h$ as implied by minimisation of loop corrections to the above potential. 

One can then calculate lifetime of the vacuum numerically solving the resulting equation of motion of the scalar field to obtain the bounce solution, or deduce a lower bound on the lifetime by using an analytical result known for quartic potential (with constant $\lambda=\lambda_c$) \cite{Lee:1985uv} to obtain the action $S=\frac{8\pi^2}{3|\lambda_c|}$. For comparison of the two methods see \cite{Lalak:2014qua}. Since $\lambda_c$ has to be negative for this solution to exist and lifetime of the vacuum depends on the action exponentially $\tau\propto e^{S}$ its easy to see that lower bound of lifetime corresponds to minimum of $\lambda$. Now our simplified lifetime reads
\begin{equation}\label{simplelifetime}
\frac{\tau}{T_U}=\frac{1}{\Lambda^4 T_U^4}e^{\frac{8\,\pi^2}{3 \,}\frac{1}{|e^{4\Gamma(\Lambda)} \lambda_c(\Lambda)|}}\, ,
\end{equation}
where $\Lambda$ is the energy scale a which $\lambda$ achieves its minimum.
This simplified result is obviously gauge dependent since at one loop in Fermi gauge we have \cite{DiLuzio:2014bua}
\begin{equation}
\gamma=\frac{1}{16 \pi^2}\left(
\frac{9}{4}g_2^2  +\frac{9}{20}g^2_1-3y^2_t-3y^2_b-y^2_\tau +\frac{3}{20}\zeta_B g^2_1+\frac{3}{4}\zeta_W g_2^2
 \right)
\end{equation}
where $\zeta_B$ and $\zeta_W$ are the gauge fixing parameters.
This problem occurs only due to the fact that we did not include the field renormalization $Z_{h}(h)=e^{\Gamma(h)}$ which appears in the kinetic term.
In our Lagrangian we always have renormalised fields $Z_{h}(h)h$ so we can use a redefined field $\phi=Z_{h}(h)h$, as the action is stationary so it is not changed by this redefinition. This can be also verified numerically, however than one has to remember that $Z_{h}$ is a function of $h$ and derivatives in the kinetic term act on it accordingly.

\begin{figure}[ht]
\centering
\includegraphics[scale=1.0]{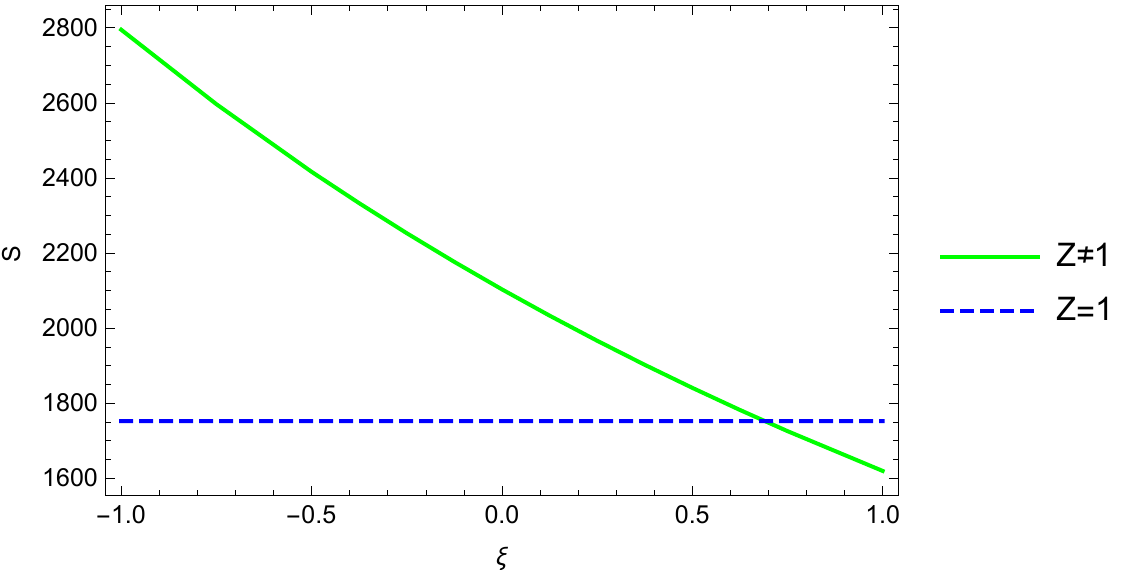} 
\caption{Gauge dependence of the bounce action calculated with different choices of the effective action. The solid line corresponds to the effective action which includes both running of the couplings and field renormalization. The dashed line correspond to the tree-level effective potential with running couplings treated as a function of the redefined field $\phi=Z_{h}(h)h$.} 
\label{SodXi}
\end{figure}

Action calculated using redefined field $\phi$ of course corresponds to using potential (\ref{simplepotential}) with $\Gamma=0$ and it is manifestly gauge independent. Since $Z_h$ is a positive multiplicative factor it does not change the conclusions on the absolute stability of the potential but can significantly change the border between metastability and instability. 
It is clear that using potential with field renormalisation results in more shallow true minimum and hence longer lifetime.
Gauge dependence of the bounce action calculated with different choices of the effective action is shown in the Figure~\ref{SodXi}. 
The dashed lines correspond to the tree-level effective potential with running couplings and field renonrmalization. The solid line corresponds to the tree-level effective potential with running couplings treated as a function of the redefined field $\phi=Z_{h(h)}h$. It is very important notice that using Landau gauge ($\xi = 0$) predicts an incorrect action of the bounce.

Table~\ref{lifetimetable} shows how the resulting lifetime changes in case of the Standard Model using the Landau gauge. We can see that incorrectly including $Z_h$ in the potential leads to lifetime longer by a factor of more than $10^{100}$. We would like to point out that our results in \cite{Lalak:2014qua} have already been supplemented with this correction despite the fact that we did not point this out explicitly in that paper.
The differences between various choices of the effective action are illustrated in the Table~\ref{lifetimetable}. 
\begin{table}[ht] \label{lifetimetable}
\begin{center}
\begin{tabular}{|c|c|c|} 
\hline 
Potential type&Bounce action $S_B$&lifetime $\tau$\\ \hline \hline
Tree level, RGE improved, $Z = 1$&$S_B = 1764$&$ \tau=1.09*10^{529}$ \\ \hline
Tree level, RGE improved, Landau gauge &$S_B = 2111$ & $\tau=2.55*10^{676}$ \\ \hline
\end{tabular}
\caption{Lifetime of the EW vacuum for different methods of finding the bounce solution}
\end{center}
\label{lifetimetable}
\end{table}%

\section{New interactions and Standard Model  phase diagram}
\begin{figure}[ht]
\begin{minipage}[t]{0.29\linewidth} 
\centering
\includegraphics[scale=0.8]{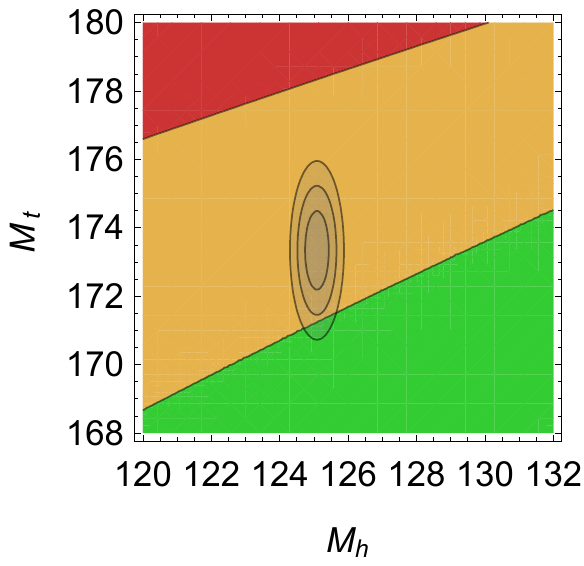} 
\end{minipage}
\hspace{0.5cm}
\begin{minipage}[t]{0.33\linewidth}
\centering 
\includegraphics[scale=0.8]{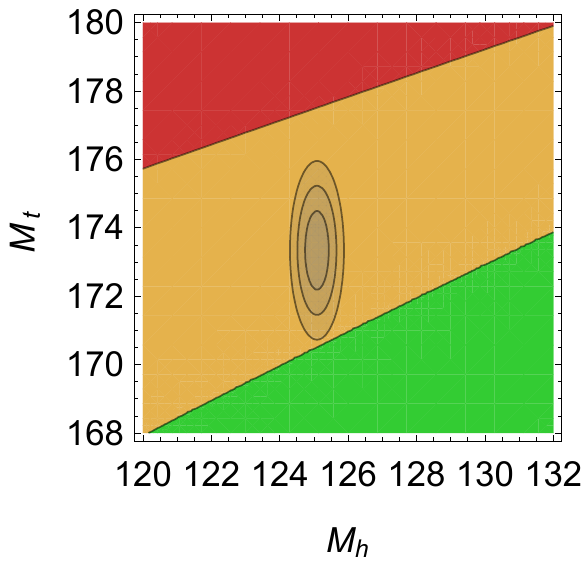}  
\end{minipage}
\begin{minipage}[t]{0.31\linewidth}
\centering 
\includegraphics[scale=0.8]{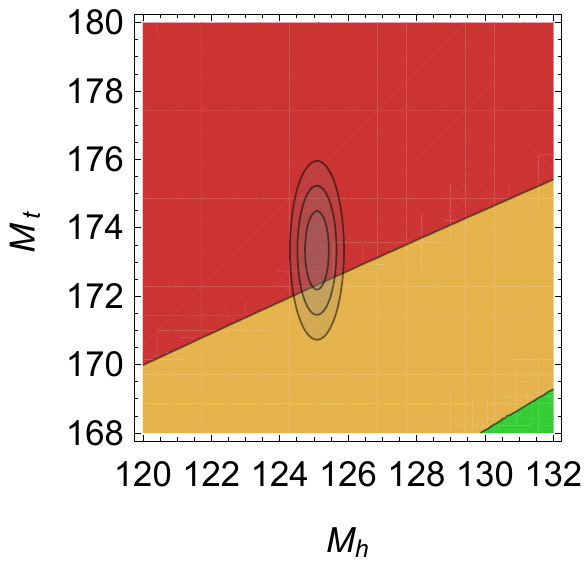}  
\end{minipage}
\caption{
Standard Model phase diagram (left panel), the same diagram after including new operators $\lambda_6(M_p)=-1/2$ and  $\lambda_8(M_p)=1$ (middle panel) and $\lambda_6(M_p)=-1$ and  $\lambda_8(M_p)=1/2$ (right panel). The green region corresponds to absolute stability, the red region to instability, and the yellow region to metastability.
\label{phasediagram}
}
\end{figure}

In what follows we will assume the Lagrangian of the Standard Model augmented by two higher dimensional operators proportional to $|H|^6$ and $|H|^8$, where $H$ is the Higgs doublet. They are suppressed by a large mass scale $M$ to an appropriate power. Being interested only in the direction $H=(\phi / \sqrt{2},0)$, we obtain a potential of the form:
\begin{equation}\label{potential}
V=-\frac{m^2}{2} \phi^2 +\frac{\lambda}{4} \phi^4 + \frac{\lambda_6}{6!}\frac{\phi^6}{ M^2} + \frac{\lambda_8}{8!}  \frac{\phi^8}{M^4}.
\end{equation}
One-loop beta functions of new couplings take the form
\begin{eqnarray} \label{betafunctions}
16\pi^2\beta_{\lambda_6} &=& \frac{10}{7}\lambda_8\frac{m^2}{M^2}+18\lambda_6 6\lambda-6\lambda_6\left(\frac{9}{4}g_2^2+\frac{9}{20}g_1^2-3y_t^2\right), \\ 
16\pi^2\beta_{\lambda_8} &=& \frac{7}{5}28\lambda_6^2+30\lambda_8 6\lambda-8\lambda_8\left(\frac{9}{4}g_2^2+\frac{9}{20}g_1^2-3y_t^2\right), \nonumber
\end{eqnarray}
which agrees with \cite{Jenkins:2013zja}.
To illustrate effects of new nonrenormalisable operators on Standard Model vacuum stability we show in Figure~\ref{phasediagram} the well known Standard Model phase diagram (see for example \cite{Buttazzo:2013uya}) and the same diagram after inluding new operators, respectively  $\lambda_6(M_p)=-1/2, -1$ and  $\lambda_8(M_p)=1, 1/2$.
Above plots clearly show that nonrenormalisable interactions supressed  by the Planck mass can drastically change the SM phase diagram, by pushing electroweak vacuum towards the instability region. 

\section{Lowering the magnitude of the suppression scale}
\begin{figure}[ht]
\begin{minipage}[t]{0.47\linewidth}
\centering
\includegraphics[scale=0.6]{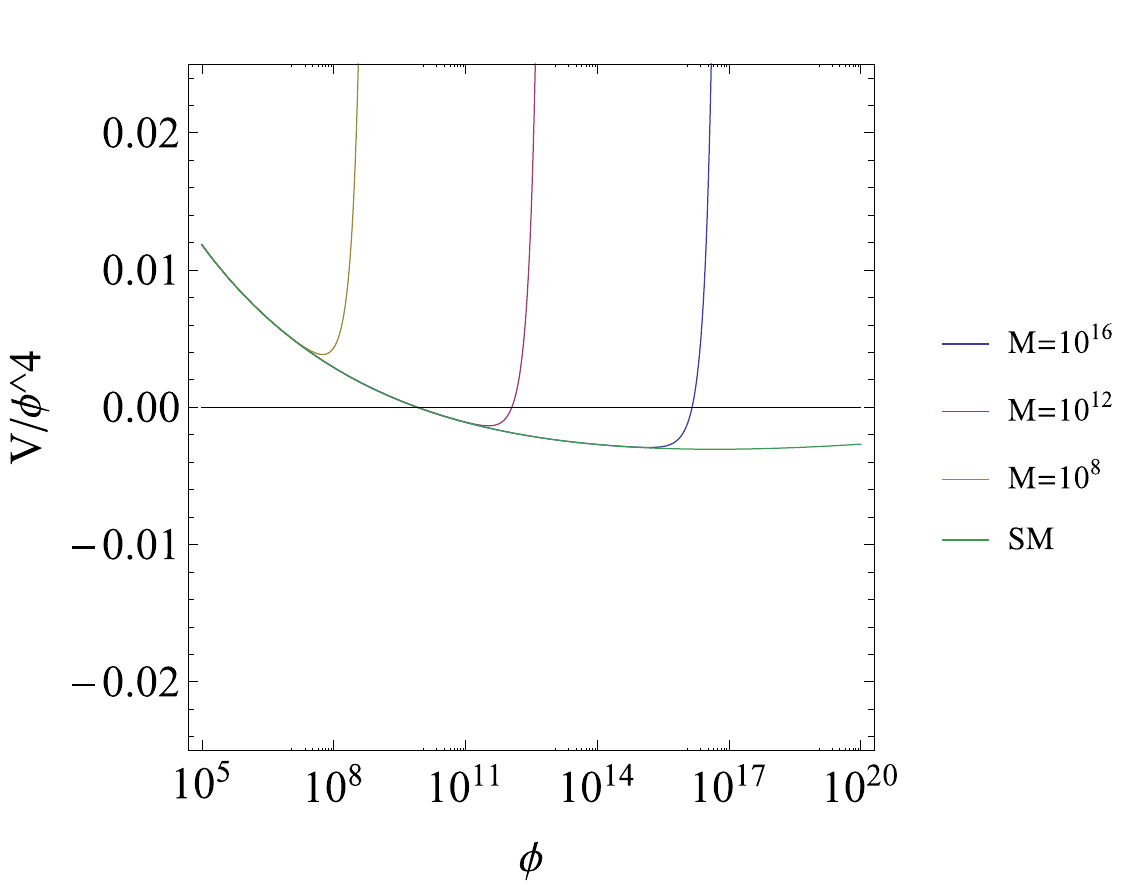}
\end{minipage} 
\begin{minipage}[t]{0.47\linewidth}
\centering
\includegraphics[scale=0.6]{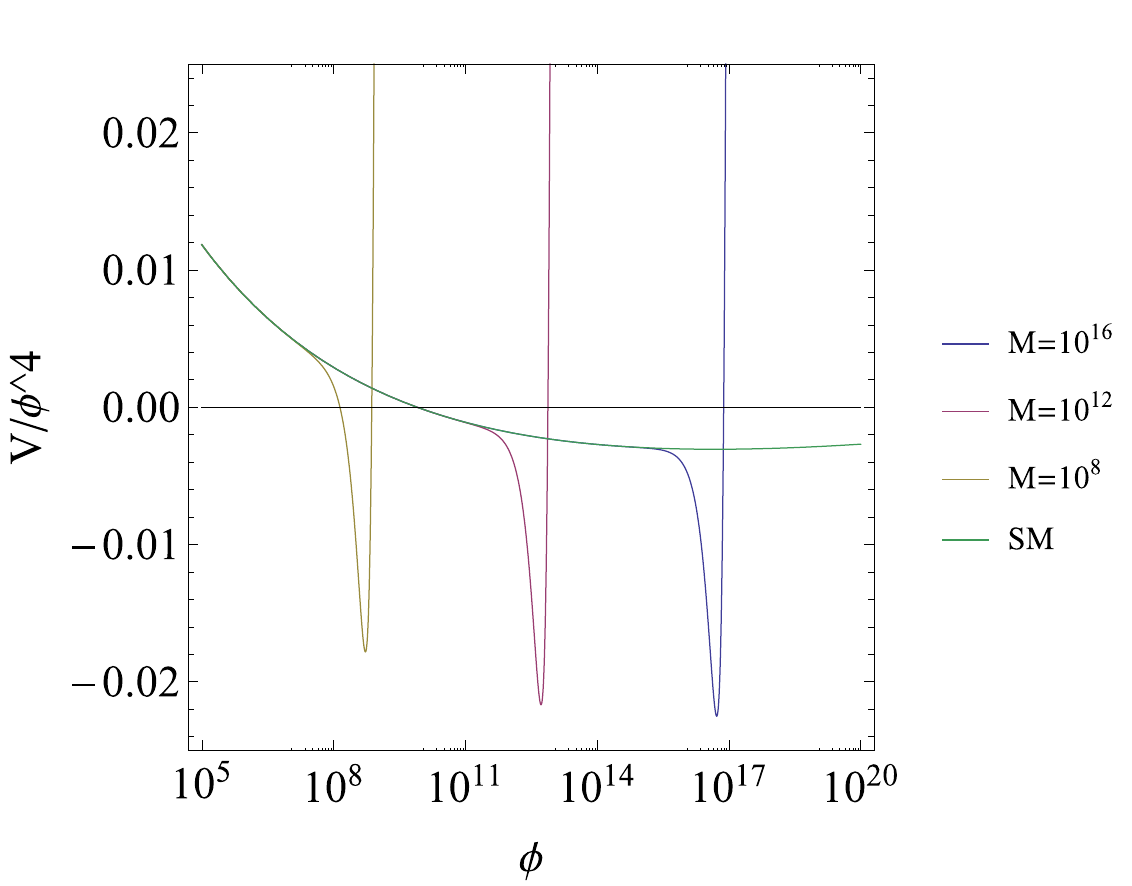}
\end{minipage} 
\caption{Scale dependence of $\lambda_{eff}/4 = V/\phi^4$ with $\lambda_6=\lambda_8=1$ (left panel) and $-\lambda_6=\lambda_8=1$ (right panel) for different values of suppression scale $M$. The lifetimes corresponding to suppression scales $M=10^{8},10^{12},10^{16}$ are respectively, $\log_{10}(\frac{\tau}{T_U})=\infty,1302,581$ (left panel) and $\log_{10}(\frac{\tau}{T_U})=-45,-90,-110$ (right panel) while for the Standard Model $\log_{10}(\frac{\tau}{T_U})=540$.
\label{scaleplots1}
}
\end{figure}

It is interesting to see how lowering the suppression scale $M$ in (2.1) changes our results.  To analyse this problem qualitatively it is enough to use the analytical approximation (\ref{simplelifetime}). When nonrenormalisable operators are positive, lowering the suppression scale $M$ corresponds simply to making the potential positive not far above $M$. The action (the exponent in (\ref{simplelifetime})) increases because the position of the minimum of $\lambda_{eff}$ shifts towards smaller energy scales and the value of $|\lambda_{eff}|$ decreases, which is shown in left hand side panel of Figure~\ref{scaleplots1}. 
  
In the case with positive $\lambda_8$ and negative $\lambda_6$ this dependence is smaller.
The new minimum is deeper and changing the scale changes $\lambda_{eff}$ by a small fraction of its value which means the resulting lifetimes are much less scale dependent, as shown in the right hand side panel of Figure~\ref{scaleplots1}. 
 
The last possibility is a potential unbounded from below ($\lambda_8<0$) which corresponds to quickly decaying solutions ( see \cite{Lalak:2014qua}).

\section{Summary}
In this note we have presented a map of the vacua in the SM extended by nonrenormalisable scalar couplings, taking into account the running of the new couplings and going beyond the standard assumptions taken when calculating the lifetime of the metastable vacuum. As a preliminary step  we  have discussed the issue of gauge dependence of the perturbative determination of the tunnelling rate and have argued how this rate can be made gauge independent at the leading nontrivial order of the RGE improved effective action.

%
In general, we confirm that it is relatively easy to destabilise the  SM with the help of the Planck scale suppressed scalar operators. While it is possible to stabilise the SM  by adding such higher dimensional interactions and taking into account running of the new couplings, the generic effect is shortening the lifetime and hence further destabilisation of the SM electroweak vaccuum. It has been demonstrated that absolute stability can be achieved by lowering the suppression scale of higher order operators while picking up such combinations of new couplings, which do not generate new deep minima in the potential. Our results also show the dependence of the lifetime of the electroweak minimum on the magnitude of the new couplings. 
\begin{center}
{\bf Acknowledgements}
\end{center}
 This work was supported by the Foundation for
Polish Science International PhD Projects Programme co-financed by the EU
European Regional Development Fund. This work has been supported by National Science Centre 
under  research grants DEC-2012/04/A/ST2/00099 and  DEC-2014/13/N/ST2/02712.

\end{document}